\documentclass[twocolumn,aps]{revtex4}

\usepackage{graphicx}
\usepackage{dcolumn}
\usepackage{bm}
\usepackage{epsfig}
\usepackage[usenames, dvipsnames]{color} 

\begin{document}

\title{First principles search for $n$-type oxide, nitride, and sulfide thermoelectrics}

\author{Kevin F. Garrity}
\email{kevin.garrity@nist.gov}
\affiliation{%
Material Measurement Laboratory, National Institute of Standards and Technology, Gaithersburg MD, 20899
}%

\date{\today}

\begin{abstract}
Oxides have many potentially desirable characteristics for
thermoelectric applications, including low cost and stability at high
temperatures, but thus far there are few known high $zT$ $n$-type
oxide thermoelectrics.  In this work, we use high-throughput first
principles calculations to screen transition metal oxides, nitrides,
and sulfides for candidate materials with high power factors and low
thermal conductivity.  We find a variety of promising materials, and
we investigate these materials in detail in order to understand the
mechanisms that cause them to have high power factors.  These
materials all combine a high density of states near the Fermi level
with dispersive bands, reducing the trade-off between the Seebeck
coefficient and the electrical conductivity, but they do so for
several different reasons.  In addition, our calculations indicate
that many of our candidate materials have low thermal conductivity.
\end{abstract}

\maketitle

\section{\label{intro}Introduction}

The need for clean efficient power generation has led to a renewed
interest in thermoelectric materials, which can directly convert a
temperature gradient into electrical power.  Thermoelectrics can
take advantage of a variety of heat sources, including solar or waste
heat, to cleanly generate
electricity\cite{oxide_thermo_review1,oxide_thermo_review2,oxide_thermo_review3,thermo_review,thermo_applications}.
Conversely, they could be used in cooling applications via the Peltier
effect.  There has been an extensive effort over recent years to
discover and optimize materials with high $zT$, a dimensionless
thermoelectric figure of merit.  While there has been significant
progress in this area, existing materials have not yet managed to
provide a combination of high $zT$, low materials cost, and high
durability that would result in widespread adoption.  Much of the
research on thermoelectrics has focused on high mobility
semiconductors with small band gaps.  Unfortunately, many of the most
promising candidate materials have practical concerns (cost, toxicity,
stability) which have thus far limited their use in applications\cite{thermo_data_review}.

In this work, we focus on the less explored group of wide band gap
transition metal oxides, as well as related nitrides and
sulfides. While oxides are not usually thought of as promising for
thermoelectric applications, due to their typically low mobilities, the
discovery of good thermoelectric performance in $p$-type Na$_x$CoO$_2$
and other layered Co-based materials resulted in an increased interest
in this class of materials\cite{oxide_thermo_review1,oxide_thermo_review2,oxide_thermo_review3,naxcoo2_expt}.  $n$-type materials
such as ZnO and SrTiO$_3$ have also displayed high power factors, but
their $zT$ values have thus far been only moderate, due to high
thermal conductivities\cite{zno_expt,srtio3_expt, srtio3_expt2}.
Despite limited success thus far, oxides provide many potential
advantages as thermoelectrics: 1) high thermal and chemical stability
in air, 2) chemical versatility, allowing
for extensive substitutions and doping, 3) low thermal conductivity,
and 4) low cost materials and
processing\cite{thermo_data_review, oxide_thermo_review1,oxide_thermo_review2,oxide_thermo_review3,
  oxide_thermal_cond}.  Thus far, much of the experimental work on
oxide thermoelectrics has focused on a relatively small number of oxides,
mostly binaries and perovskites, leaving open the possibility that
better oxide thermoelectrics exist.

In this work, we use high-throughput density functional theory (DFT) 
calculations\cite{compmatsci,ceder,ceder2,wolverton,datamine,cluster,evolutionary,curtarolo2,curtarolo,zunger_abc,database,ht_highway}
to identify promising $n$-type thermoelectric oxides and related
materials from the Inorganic Crystal Structure Database
(ICSD)\cite{icsd}. The large amount of work required to synthesize,
optimize, and measure thermoelectrics experimentally make this type of
theoretical screening of candidate materials particularly desirable.
Similar techniques have been used successfully to study the
thermoelectric behavior of a variety of materials, including
oxides\cite{thermoelectric, silicides, ab_thermoelectrics, oxide_thermo_theory, thermoelectric_computation, datamine_thermoelectrics,thermoelectric_predictors, naxcoo2_theory,srtio3_theory,
  thermoelectric_theory, ktao3_theory, automated_search_te}.  While a fully first principles theoretical
calculation of $zT$ remains challenging, especially for oxides, which often have partially localized carriers, we can nevertheless screen materials
for both electronic and vibrational properties that are necessary for good thermoelectric
performance.  In this work, we perform such a screening procedure, identifying many
candidate materials with calculated thermoelectric properties that are
similar to or surpass experimentally studied $n$-type oxides.  Furthermore, we
analyze the mechanisms behind the high thermoelectric performance of
these materials, finding that they fall into a small number of groups
with similar properties.


\section{\label{methods}Methods}

\subsection{Calculating Thermoelectric Performance}

The dimensionless figure of merit for thermoelectrics can be written as
\begin{equation}
zT = \sigma S^2 T / \kappa,
\end{equation}
where $\sigma$ is the electrical conductivity, $S$ is the Seebeck
coefficient, $\kappa$ is the total thermal conductivity (electrical
plus lattice), and $T$ is the temperature. The power factor,
which determines the electrical response of a material to a
temperature gradient, is $S^2 \sigma$. 

Unfortunately, the components of $zT$ are not all easy to calculate
using first principles techniques.  Within the constant relaxation
time approximation, which is used in this work, $S$ can be calculated
from a band structure calculation without any adjustable
parameters\cite{ziman}. Within the same approximation, it is
possible to calculate $\sigma / \tau_e$, where $\tau_e$ is the
electronic relaxation time.  Unfortunately, calculating $\tau_e$ from
first principles remains
challenging\cite{silicon_electrical_cond,tio2_polaron}. This problem
is especially severe for oxides, which often display complicated
conduction mechanisms and polaronic effects at low doping and low
temperatures.  In this work, we are concerned primarily with the
opposite regime of high temperatures and high doping, where the carrier mobilities of oxides
are typically
larger\cite{oxide_thermo_review1,oxide_thermo_review2,oxide_thermo_review3}.
Because we are comparing materials which are chemically similar, we
expect them to have broadly similar electron scattering mechanisms.
Therefore, we will use the quantity $S^2 \sigma / \tau_e$ to rank our
candidate materials for suitability as thermoelectrics.  This
estimate, which has been used in many previous
works\cite{thermoelectric_theory, silicides, non_parabolic}, should be
sufficient to at least screen materials for those with band structures
that are promising for thermoelectric applications, even if
determining the final ranking of materials will require experimental
input.

For reference, first
principles techniques can reproduce the thermoelectric properties of
SrTiO$_3$ with $\tau_e \approx 4$ fs at room
temperature\cite{srtio3_theory}, a typical value for oxides, but some
high mobility oxides like ZnO have much longer scattering
times\cite{zno_theory}. All wide band materials have to be doped in order to be used as
thermoelectrics.  In this work, we use the rigid band filling model to
estimate the effects of doping, and we rank materials by $S^2 \sigma /
\tau_e$ at their optimum doping.

After identifying materials with promising band structures, we perform
more computationally expensive phonon calculations for a limited number of candidate materials to estimate the lattice thermal
conductivity, which is the dominant contribution to the thermal conductivity for most
thermoelectrics, as described in Sec.~\ref{thermal}.

\subsection{Band Structure Calculations}

All of our calculations are based on DFT calculations\cite{hk,ks}, as
implemented in QUANTUM ESPRESSO\cite{QE} and using the GBRV
high-throughput ultrasoft pseudopotential library\cite{gbrv}.  We use
a plane wave cutoff of 40 Ryd for band structure calculations and
45-50 Ryd for phonon calculations. For Brillouin zone integration, we
use a $\Gamma$-centered grid with a density of 1500 k-points per atom.

We use the PBEsol exchange-correlation functional\cite{pbesol}, which
provides more accurate lattice constants and phonon frequencies than
other GGA functionals.  We use the DFT+U technique\cite{ldaplusU,
  ldaplusU_simplified, dftU_simp}, with a U value of 3 eV for
transition metal $d$-states\cite{dftU_values}, when calculating band
structure related quantities. We find that for most materials this
correction has a relatively minor effect beyond increasing the band
gap, and larger gaps have no direct effect on thermoelectric
performance as long as the gap is already large enough to avoid
significant thermal carrier excitation. We perform phonon calculations
using DFT perturbation theory\cite{dft-pt} without the +U correction.

Our main results are done on fully relaxed structures with initial
coordinates from the ICSD.  We use PYMATGEN\cite{pymatgen} to
manipulate files from the ICSD to setup the initial structures for
relaxation. Because calculations are run at a fixed number of plane
waves, changes in the unit cell during relaxation can effectively
modify the basis set. To ensure consistency between the basis set and
the final structure, we run each relaxation three times, with a force
convergence tolerance of 0.001 Ry/Bohr, an energy tolerance of $1
\times 10^{-4}$ Ry, and a stress tolerance of 0.5 Kbar. For phonon calculations, we decrease the force
tolerance to $5 \times 10^{-5}$ Ry/Bohr.  The BFGS algorithm as implemented in QUANTUM ESPRESSO was used for relaxations.

We use maximally localized Wannier functions as implemented in
WANNIER90\cite{mlwf_orig, mlwf_dis, wannier90} to interpolate band structures and
BOLTZWANN, the WANNIER90 transport module, to calculate the Seebeck
coefficient and conductivity under the relaxation time
approximation\cite{wann_interp, boltzwann}.  The use of Wannier
interpolation allows us to perform accurate calculations of
thermoelectric quantities starting from relatively sparse first
principles k-point grids, which we then interpolate to a k-point
spacing of 0.02 \AA$^{-1}$. This density is about ten times as dense as
the first principles calculation along each direction in k-space. The use of Wannier functions also allows
us to calculate band structure derivatives analytically, which
accurately treats degenerate points in the Brillouin zone.

In order to use Wannier interpolation for this work, we had to
develop a procedure for automating the construction of localized
Wannier functions. Because we are interested in the properties of both
valence and conduction states, we normally include all possible
orbitals which could contribute to states near the Fermi level (see
supplemental materials for list). This is in contrast to many applications of
Wannier functions, which are concerned with either only the occupied
bands or only a localized subspace of bands (e.g. d-orbitals).  In
these cases, the Wannier functions extend over several atoms and may
be sensitive to the details of the localization procedure. For our
application, we include all the relevant orbitals, which results in
Wannier functions that are atomic-like and strongly localized, even
before the iterative localization procedure, making the final result
more robust.  To calculate the Wannier functions, we use an inner
'frozen' window default of 4.5 eV above and below the
conduction/valence band edges in order to ensure an accurate interpolation of
the band structure. In testing, our calculated thermoelectric
properties are insensitive to minor variations in this window.

While the Wannierization procedure we outlined above is
relatively robust, there are a few situations that can result in
failures in the Wannierization, which are identified by
monitoring the spread of the Wannier functions.  First, if there are
semicore states that were excluded from the Wannierization that
overlap in energy with the valence states, it
will be necessary to include those states in the valence. Second,
sometimes there are problems including orbitals with high energy (e.g. Sr
$d$-states in SrTiO$_3$), as these states can become difficult to
disentangle from the free electron-like bands when their energy becomes
too high.  In both of these cases, we simply adjust the orbitals that
we include in the Wannierization procedure by hand to fix these problems.
Another issue can arise if the 'frozen' window overlaps with free
electron-like bands.  This can be fixed by adjusting this window
downward to avoid overlap.  We encountered all of these problematic cases only rarely,
and we adjust for them when necessary.

One potential drawback of the Wannierization approach is the necessity of
including a large number of empty bands in a non-self-consistent DFT
calculation, in order to construct well-localized conduction band
Wannier functions. However, these extra bands are only required on the
the sparse k-point grid, and in practice the computational cost of this
step is smaller than the initial structural relaxation.

\subsection{\label{thermal}Thermal conductivity}

For typical thermoelectrics, the thermal conductivity is dominated by
the lattice thermal conductivity ($\kappa_l$). First principles
calculations of the thermal conductivity have been shown to be
accurate for a wide variety of
materials\cite{lattice_thermal_cond,si_heat,thermal_compound_semiconductors,
  phonon_lifetimes, hh_thermal}. Unfortunately, these calculations
require the anharmonic force constants, which are too computationally
expensive to use as an initial screening tool for high-throughput
calculations, especially as many of the materials we consider have
large unit cells with relatively low symmetry.

There have been various recent attempts to model the lattice thermal
conductivity without performing a full calculation of the anharmonic
force constants\cite{hh_thermal, model_thermal, bayesian_thermo,
  ht_thermal_cond, thermoelectric_predictors, thermalcond_trends}.
Yan \textit{et. al.} use a Debye-Callaway model with a constant
Gruneisen
parameter\cite{thermoelectric_predictors,thermalcond_trends}.  Toher
\textit{et. al.}\cite{ht_thermal_cond} demonstrated that a modeled the
Debye temperature and the Gruneisen parameter, combined using the
Slack model\cite{slack, slack_2} for thermal conductivity, is useful as a
screening method for thermal conductivity.  Another screening method
by Bjerg \textit{et. al.}\cite{model_thermal, thermal_quasi_harmonic} incorporates aspects of
the first principles phonon band structure to approximate the lattice
thermal conductivity.  

In this work, we want a method which is both computationally feasible
to apply to a few dozen compounds to use as a secondary screening
procedure, and accurate enough to provide a reasonable ordering of
compounds to consider for further study. We employ a method similar to
the model in Bjerg \textit{et. al.}, where the Gruneisen parameter ($\gamma$) and
the Debye temperature ($\Theta_D$)are calculated from the first principles phonon
dispersion
\begin{eqnarray}\label{eq:debye}
\Theta_D &=& n^{-1/3} \sqrt{\frac{5 \hbar^2}{3 k_B^2}\frac{\int_0^\infty \omega^2 g(\omega) d\omega}{\int_0^\infty g(\omega) d\omega}} \\
\gamma^2 &=& \frac{\sum_i \int \frac{d\bf{q}}{8 \pi^3}  \gamma_{i\bf{q}}^2C_{i\bf{q}}}{\sum_i \int \frac{d\bf{q}}{8 \pi^3}  C_{i\bf{q}}} \\
\gamma_{i\bf{q}} &=& -\frac{V}{\omega_{i\bf{q}}}\frac{\partial \omega_{i\bf{q}}}{\partial V} 
\end{eqnarray}
where $n$ is the number of atoms per unit cell, $\omega_{i\bf{q}}$ is
the angular frequency of phonon mode $i$ at $q$-point $\bf{q}$,
$g(\omega)$ is the phonon density of states, $\gamma_{i\bf{q}}$ is the
mode Gruneisen parameter, $C_{i \bf{q}}$ is the mode specific heat,
and $V$ is the volume.  The sum for the Gruneisen parameter is only
performed over modes with $\hbar \omega_{i \bf{q}} < k_B \Theta_D$.
As per the discussion in
Refs. \onlinecite{model_thermal,thermal_quasi_harmonic}, we square
$\gamma_{i\bf{q}}$ to avoid cancellation between positive and negative
anharmonicity when calculating $\gamma$.

Using the Debye temperature and Gruneisen parameter calculated above, we then insert them into the Slack model \cite{model_thermal, slack, slack1,slack2,slack_2}:
\begin{eqnarray}\label{eq:kappa_l}
\kappa_l(T) = \frac{0.849 \times 3 \sqrt[3]{4}}{20 \pi^3 (1 - 0.514 \gamma^{-1} + 0.228 \gamma^{-2})} \nonumber \\
\times \Big(\frac{k_B \Theta_D}{\hbar}\Big)^2 \; \frac{k_B M V^{\frac{1}{3}}}{\hbar \gamma^2} \frac{\Theta_D}{T} 
\end{eqnarray}

where $M$ is the average atomic mass.  We find that the Debye
temperature and Gruneisen parameter used in this way contain almost
all of the information of the full Bjerg model, as shown in table
\ref{tab:thermal}, which presents correlations of various models and
quantities with our reference set of thermal conductivities.  In fact, as shown in table \ref{tab:thermal}, 
the quantity $\Theta_D / \gamma$ also has a high rank correlation with the reference thermal conductivities, 
although there is no computational advantage in using $\Theta_D / \gamma$ instead of Eq. \ref{eq:kappa_l}.  

The reason we do not use the full Bjerg model is
that in some cases we found difficulty in fully converging the
acoustic modes for large unit cells.  These modes depend on careful
cancellation between all of the force constants to produce modes with
zero eigenvalues at $q=\Gamma$, which is challenging to achieve
numerically.  This cancellation can be enforced at $\Gamma$ by
using the acoustic sum rule to modify the force constant matrix in various ways, but we sometimes
found results which depend on how the rule was enforced.  Therefore, we
opted to use a more computationally robust procedure appropriate for a
high-throughput study by using the Slack model instead of the Bjerg
model.

We find empirically in testing that using this combination of the Slack model
with the Bjerg definition of the Debye temperature and Gruneisen
parameter overestimates the thermal conductivity, so we report 70\% of
the model value, which improves the quantitative accuracy in our
testing but makes no difference in a ranking of compounds for those
with the lowest thermal conductivity. For materials with unstable
phonon modes, we cannot calculate a Gruneisen parameter in a
meaningful way using purely harmonic calculations, as the unstable
modes must be stabilized anharmonically at finite temperature.
Therefore, we do not estimate a thermal conductivity for those compounds.
In practice, we expect many materials with unstable modes
at zero temperature to have low thermal conductivity due to anharmonic
interactions, so the observation of unstable modes is already a useful indicator of anharmonicity. 
We present our calculated thermal conductivities at 300 K, even
though we expect these materials to be used at higher temperatures,
where the thermal conductivity will be lower.

\begin{figure}
\includegraphics[width=3.5in]{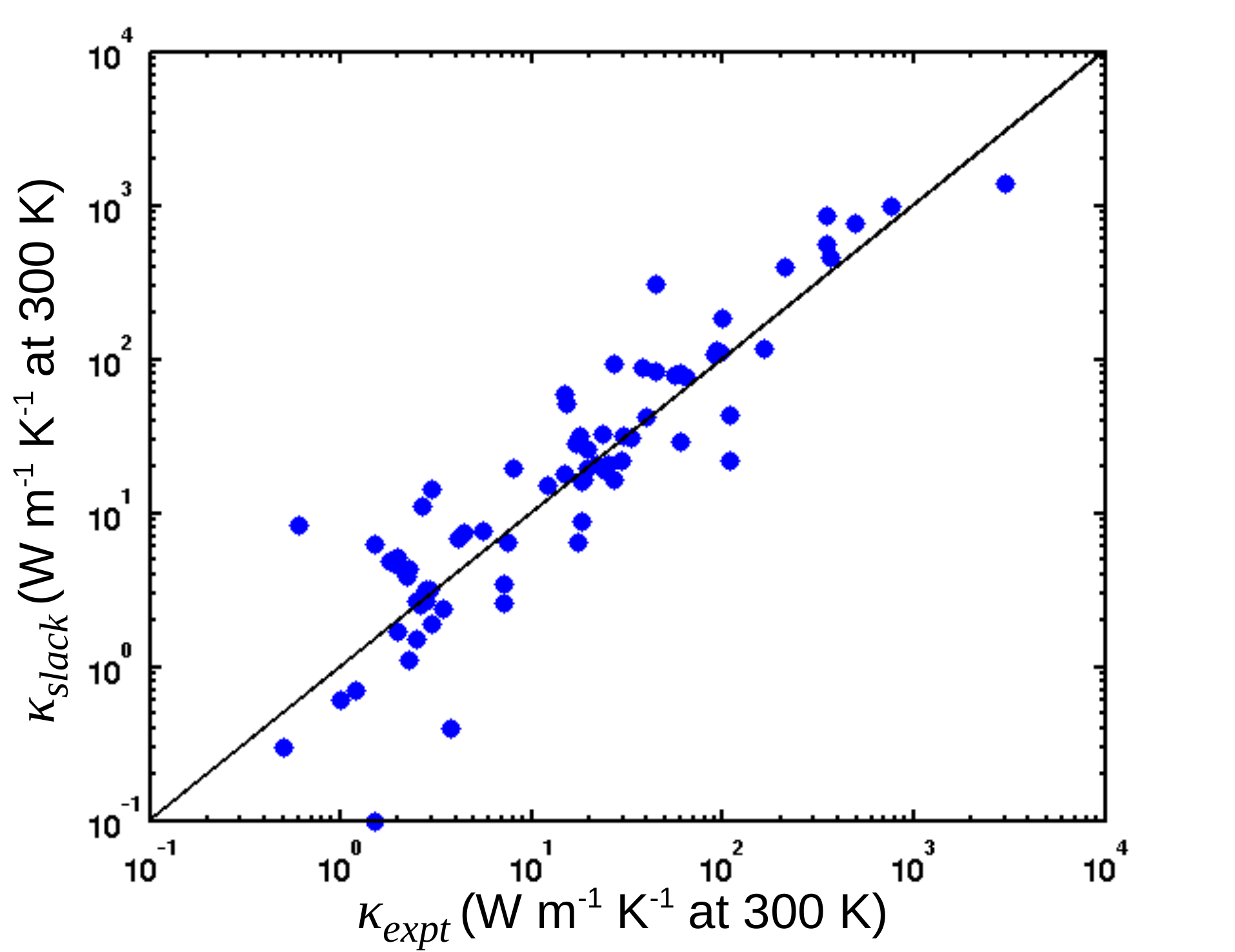}
\caption{\label{fig:kappa} Comparison of reference experimental and first principles thermal conductivities (x-axis) and
  the Slack model (y-axis).}
\end{figure}

\begin{table}
\caption{\label{tab:thermal} Correlations of various quantities with the reference thermal conductivities at 300 K (see Supplementary Materials for list). $\kappa_{Slack}$ is the model used in this work (see Eq. \ref{eq:kappa_l}) ; Bjerg refers to the full model of Ref. \cite{model_thermal}.  Pearson and Spearman refer to the standard Pearson correlation and the Spearman rank correlation, respectively. The first two columns include all materials, the next two are limited to materials with $\kappa_l < 50 $ W m$^{-1}$K$^{-1}$ at 300 K.}
\begin{ruledtabular}
\begin{tabular}{lcccccccccc}
Quantity & Pearson & Spearman & Pearson & Spearman \\
         &      &      &  Low $\kappa_l$& Low $\kappa_l$ \\
\hline
$\kappa_{Slack}$          & 0.83 & 0.91 & 0.65 & 0.83 \\
Bjerg\cite{model_thermal} & 0.93 & 0.92 & 0.69 & 0.88 \\
$\Theta_D$                & 0.74 & 0.82 & 0.43 & 0.71 \\
$1/\gamma$                & 0.39 & 0.75 & 0.66 & 0.60  \\
$\Theta_D / \gamma$       & 0.71 & 0.89 & 0.60 & 0.83 \\
\end{tabular}
\end{ruledtabular}
\end{table}

We establish the validity of this method for screening the thermal
conductivity by comparing the model with the experimental thermal
conductivities of a variety simple binary semiconductors, as well as
the first principles thermal conductivities for a variety of
half-Heusler compounds \cite{ht_thermal_cond, slackbook, hh_thermal}.  In addition, we compare with
experimental thermal conductivities for a few additional
oxides\cite{thermal_perovskites}.  The results are shown graphically
in Fig.~\ref{fig:kappa} and correlations are given in table \ref{tab:thermal}, see the supplementary materials for details.
We find that our chosen method is sufficient for screening materials
for those likely to have low thermal conductivity.  As shown in table \ref{tab:thermal}, the Spearman rank
correlation\cite{ht_thermal_cond} between the reference and modeled
thermal conductivities for the entire test set is 0.91, indicating we
are able to identify promising materials.  If we limit the dataset to
materials with $\kappa_l < 50$ W m$^{-1}$K$^{-1}$ at 300 K, a more
realistic range for complex oxides, the rank correlation for the full
model drops to 0.83, which is still reasonable for selecting materials
to study further.  We also note that when considering the entire dataset, the Debye temperature (see Eq. \ref{eq:debye}) alone has a
rank correlation of 0.82 with the thermal conductivity, and we make it as
an initial screening tool, as it is less computationally expensive
than the full model.  However, directly calculating the Gruneisen
parameter, rather than estimating it\cite{ht_thermal_cond, thermoelectric_predictors, thermalcond_trends}, significantly increases the
accuracy of the model.  Finally, experimental thermal conductivities
are sensitive to many factors beyond the scope of this work, including
defects and grain boundaries, which both makes comparisons with
experiments difficult but increases the possibility of engineering
materials to have lower thermal conductivities.

\subsection{\label{meff}Effective Masses}

In order to understand the conductivity and Seebeck coefficient, we
consider several definitions of the effective mass. Using our Wannier
interpolation, we calculate derivatives of the band structure
analytically\cite{wann_interp} at the conduction band minima to find
the effective mass tensor $(m_{ij})^{-1} = \frac{1}{\hbar^2}\frac{d^2
  E}{dk_i dk_j}$.  We will sometimes concentrate on $m_{min}$, the smallest eigenvalue of $m_{ij}$, which helps 
determine the largest value of the conductivity tensor at low temperature. 
We will also consider $m_{iso}=(m_1 m_2 m_3)^\frac{1}{3}$, 
the isotropic effective mass, where $m_1$, $m_2$, and $m_3$ are the eigenvalues of $m_{ij}$.

A related band structure descriptor we calculate is a version of the effective mass based on the density of states (DOS):\cite{mdos1,mdos2,thermoelectric_predictors}
\begin{eqnarray}\label{eq:mdos}
m_{DOS}(E) &=& \hbar^2 \sqrt[3]{\pi^4 g(E)g'(E)} \\
\label{eq:mdos2}
m_{DOS}(T,n_d) &=&  \frac{\int dE \, g(E) m_{DOS}(E)  (-\frac{df}{dE})}{dE \, g(E) (-\frac{df}{dE})}.
\end{eqnarray}
In this expression, $g(E)$ is the DOS at energy $E$, $f(E)$ is the
Fermi function, and $T$ and $n_d$ are the temperature and doping.
This definition of $m_{DOS}$ matches $m_{eff}$ for a single parabolic
band, but it is higher for non-parabolic bands or when multiple bands
contribute to the conduction.  These features allow $m_{DOS}$ to give
a good description of the Seebeck coefficient for many
materials\cite{mdos1,mdos2,thermoelectric_predictors}.

\subsection{\label{materials}Materials selection and screening procedure}

We are interested in discovering new $n$-type thermoelectric oxides,
nitrides, or sulfides.  As there are over 80,000 entries with oxygen
in the ICSD, there is a need to significantly limit our search space
before proceeding.  Previous experimental and theoretical work on
thermoelectrics has suggested that good thermoelectrics tend to have
anisotropic and non-parabolic bands and high densities of states, all
of which can be created by empty
$d$-orbitals\cite{srtio3_theory,srtio3_lowd,low_dim_thermo,lowd_bands,non_parabolic}. Furthermore,
materials with empty $d$ orbitals can usually be doped $n$-type, with
some carriers occurring naturally due to oxygen
vacancies\cite{oxide_thermo_review1,oxide_thermo_review2,oxide_thermo_review3}.
Therefore, in this work, we focus on materials containing at least one
of Y, Sc, Ti, Zr, Hf, Nb, Ta, Mo, or W as well as at least one of O,
N, or S.  In order to limit computational time, we restrict our search
set to structures with primitive unit cell volumes of less than 300 \AA$^3$.

Our screening procedure proceeds in several steps, with each step
becoming more computationally expensive. First, we calculate the band
gap of all our our starting materials (766 compounds) and eliminate the metals.  For
materials with a gap (592 materials), we calculate the Seebeck
coefficient, $S$, and electronic conductivity, $\sigma/\tau_e$. For
materials with a high power factor (191 materials), we then calculate the Debye
temperature using Eq. \ref{eq:debye}.  Finally, 
for a subset of
19 materials with high power factor and/or low Debye temperature, 
we calculate $\kappa_l^{Slack}$ with
Eq. \ref{eq:kappa_l}. Results are presented in table \ref{tab:results} with further details in the supplementary materials.

\section{\label{results}Results and Discussion}

We performed our screening procedure starting with 766 compounds from
the ICSD as discussed in Sec.~\ref{materials}, consisting of 661 oxides, 60 nitrides, and 71 sulfides (some compounds contain multiple anions).
Of that list, we find 592 materials with band gaps according to DFT+U (551 oxides, 53 nitrides, and 25 sulfides).  For these
materials, we calculate $S$ and $\sigma / \tau_e$ for a variety of temperatures and dopings.

\begin{figure}
\includegraphics[width=3.5in]{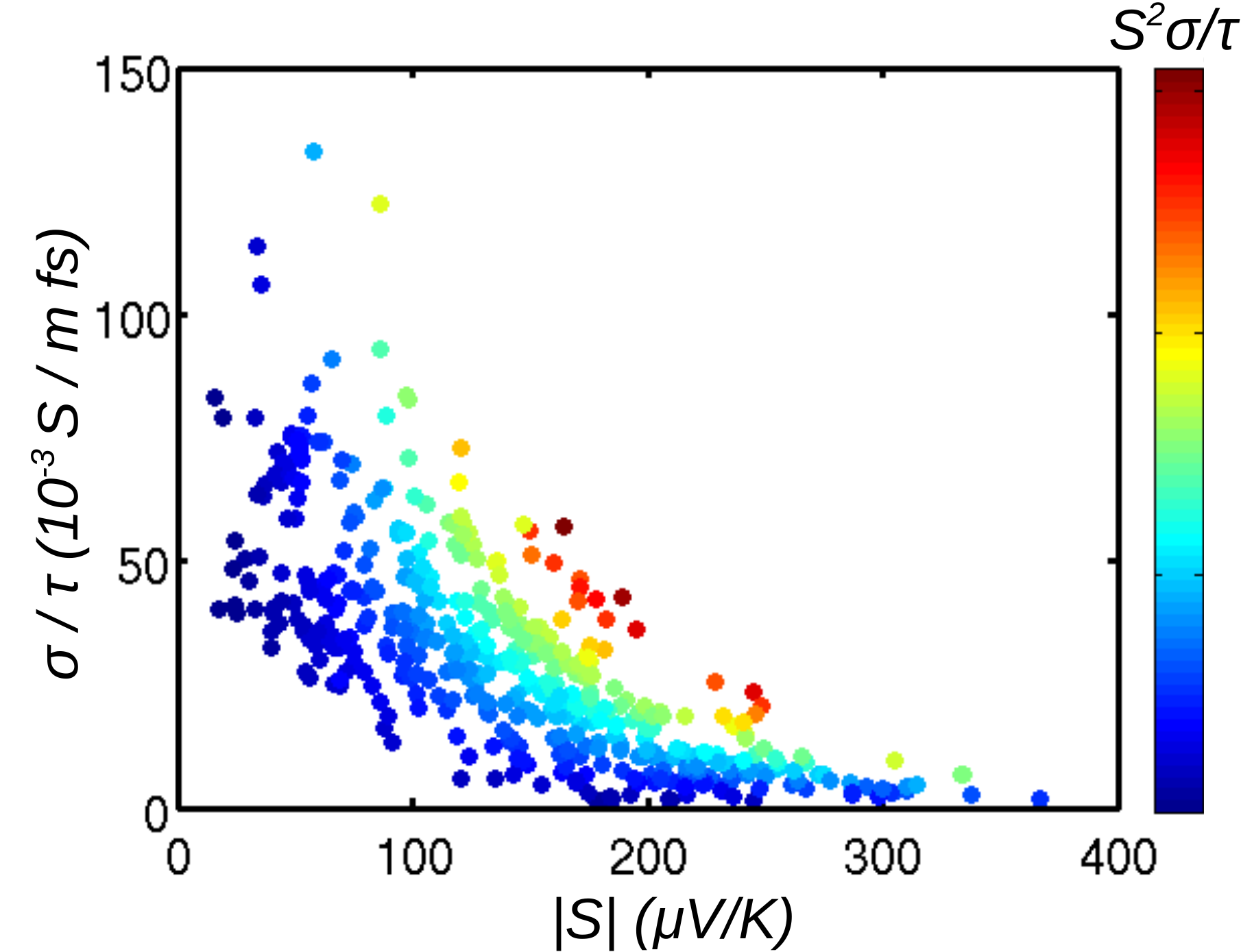}
\caption{\label{fig:all} $|S|\, (\mu V/K)$  versus
  $\sigma/\tau_e \, (10^3 \frac{S}{m \, fs})$ for the entire dataset, at T=700K and doping at $10^{21}$ cm$^{-3}$. The color scale indicates size of power factor, $S^2 \sigma / \tau_e$.}
\end{figure}

\begin{figure}
\includegraphics[width=3.5in]{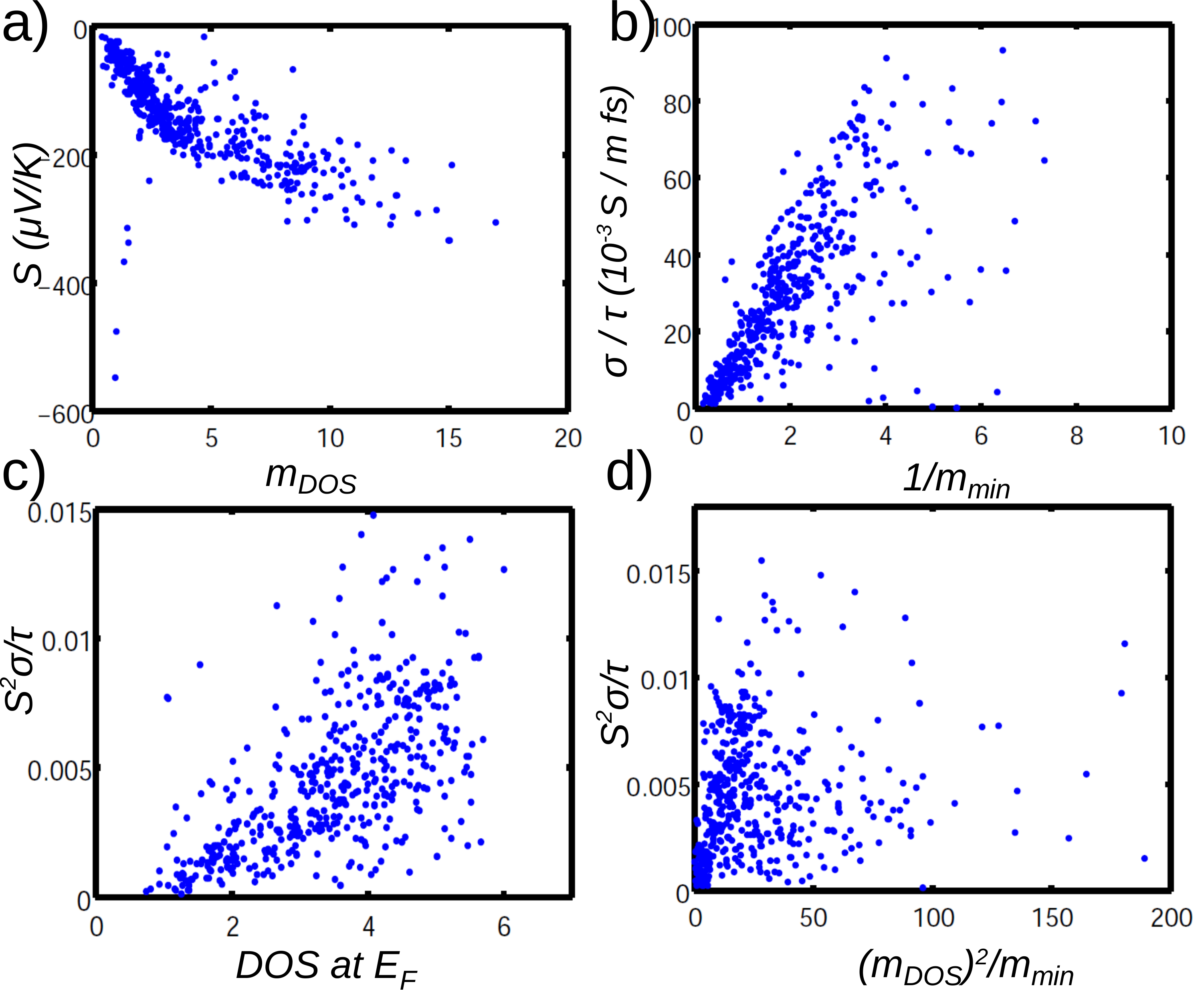}
\caption{\label{fig:comp} Thermoelectric properties of the entire
  dataset, at fixed T=700K $n_d$ = $10^{21}$ cm$^{-3}$. a) $m_{DOS}$ versus $S$, b) $1/m_{min}$ versus
  $\sigma / \tau_e$, where $m_{min}$ is the smallest component of
  $m_{ij}$, c) DOS at $E_F$ versus the power factor, and d) $m_{DOS}^2/m_{min}$ versus the power factor.}
\end{figure}

If we sort this list of candidate materials by estimated power factor,
$S^2 \sigma / \tau_e$, at 700 K and with optimized doping, we
rediscover several compounds that have previously been measured to
have good $n$-type thermoelectric properties.  For example, doped
TiO$_2$, SrTiO$_3$, KTaO$_3$, and TiS$_2$ have all been measured to
have promising power factors and show up highly in our
list\cite{oxide_thermo_review1,oxide_thermo_review2,oxide_thermo_review3,tis2,srtio3_theory,srtio3_expt,srtio3_expt2,
  ktao3_theory,ktao3_expt,ndopedtio2,tio2}. This gives us confidence
that our screening procedure is useful. In addition to these
previously measured materials, there are a variety of compounds which
have not been studied for thermoelectric applications and which may
have properties that are superior to existing materials.  In table
\ref{tab:results}, we list some of our most promising candidate
materials, including those with high $S^2 \sigma / \tau_e$ and a few
with moderate $S^2 \sigma / \tau_e$ and low $\kappa_l$.  We remove
structures that are minor distortions of other structures on the list,
that have missing atoms in the ICSD, or that are only theoretically proposaled structures; full results are presented in the
Supplementary Materials.

\begin{table*}
\caption{\label{tab:results} Thermoelectric properties of the most
  promising compounds.  The first five columns consist of the compound
  name, its space group number, the DFT+U band gap (eV), the isotropic
  effective mass.  The next several columns are the
  DOS, $m_{DOS}$ (see Eq. \ref{eq:mdos}-\ref{eq:mdos2}), $S$, and $ \sigma/\tau_e$, all evaluated at at 700 K
  and fixed $10^{21}$cm$^{-3}$ doping.  The next column is $S^2
  \sigma/\tau_e$ at the optimized doping and at 700 K, and following column is that optimized doping.  The last column is the
  modeled lattice thermal conductivity at 300 K. Materials with $-$ as $\kappa_l$ have unstable phonon modes or we were unable to converge the Gruneisen parameter.}
\begin{ruledtabular}
\begin{tabular}{lcccccccccc}
Material & Space Grp. & Band Gap & $m_{iso}$ & DOS & $m_{DOS}$ & $S$ & $ \sigma/\tau_e$ &  $S^2 \sigma/\tau_e$ & Opt. Doping & $\kappa_l$\\
         &            & (eV) &              & ($10^3$ eV$^{-1} $\AA$^{-3}) $ & & ($\mu V / K$) &10$^{-3}$ $\frac{\text{S}}{\text{m fs}}$ & 10$^3$ $\frac{W}{\text{m K}^{2} \text{fs}}$ & ($10^{21} cm^{-3}$) & $\frac{W}{m K}$ \\
\hline
CaTaAlO$_5$     & 15  & 4.0 & 2.8   & 4.08 & 5.0      & -189 & 42.9   &1.7  & 3 & $-$ \\
TiO$_2$         & 225 & 1.7 & 1.1   & 1.53 & 17.0     & -305 & 9.6    &1.7  & 8 & $-$ \\
LiNbO$_3$       & 161 & 3.5 & 2.1   & 3.92 & 7.0      & -245 & 23.4   &1.6  & 2 & 40 \\
TiO$_2$         & 136 & 2.1 & 2.2   & 3.63 & 7.5      & -250 & 20.7   &1.6  & 3 & $<$1 \\
HfS$_2$         & 164 & 1.6 & 1.6   & 5.51 & 2.7      & -165 & 56.9   &1.5  & 1 & 3.4 \\
NaNbO$_3$       & 63  & 1.8 & 3.8   & 1.89 & 0.8      & -268 & 75.6   &1.4  & 4 & $-$ \\
Ba$_2$TaInO$_6$ & 225 & 4.3 & 2.3   & 2.75 & 10.0     & -285 & 12     &1.4  & 5 & $-$ \\
YClO            & 129 & 5.1 & 1.1   & 3.20 & 6.6      & -182 & 32.2   &1.4  & 3 & 6.0 \\
LiTaO$_3$       & 161 & 3.8 & 2.9   & 5.50 & 3.4      & -204 & 36.3   &1.4  & 1 & 34 \\
Li$_2$ZrN$_2$   & 164 & 1.9 & 0.6   & 5.10 & 3.5      & -178 & 42.5   &1.4  & 2 & 19 \\
CaTiSiO$_5$     & 15  & 3.2 & 4.0   & 4.28 & 6.4      & -229 & 25.5   &1.3  & 2 & 1.1 \\
HgWO$_4$        & 15  & 2.4 & 1.5   & 4.22 & 3.5      & -171 & 46.5   &1.3  & 2 & $-$ \\
P$_2$WO$_8$     & 12  & 2.3 & 11.8  & 6.01 & 3.5      & -150 & 56.1   &1.3  & 0.7 & 6.5 \\
ZrS$_2$         & 164 & 1.1 & 1.6   & 4.88 & 3.3      & -172 & 45.1   &1.3  & 2 & 22 \\
TaPO$_5$        & 85  & 3.7 & 2.6   & 4.38 & 4.1      & -160 & 49.5   &1.3  & 2 & $-$ \\
LaTaO$_4$       & 36  & 3.4 & 2.9   & 5.13 & 3.7      & -183 & 38.1   &1.3  & 2 & 26 \\
NbTl$_3$S$_4$   & 217 & 2.3 & 0.9   & 3.59 & 8.6      & -246 & 19.2   &1.3  & 2 & $-$ \\
SrTaNO$_2$      & 140 & 0.9 & 0.7   & 3.31 & 1.8      & -109 & 122.6  &1.3 & 0.3 & $-$ \\
TiS$_2$         & 164 & 0.4 & 0.4   & 4.73 & 3.7      & -171 & 42     &1.3  & 2 & 5.3 \\
PbTiO$_3$       & 99  & 2.0 & 0.6   & 3.68 & 1.7      & -489 & 35.3   &1.2  & 7 & 16 \\
Sr$_2$TaInO$_6$ & 225 & 4.3 & 2.1   & 2.65 & 10.0     & -266 & 10.4   &1.2  & 6 & $-$ \\
SrTiO$_3$       & 140 & 1.2 & 1.3   & 3.56 & 1.6      & -165 & 39.5   &1.2  & 6 & $-$ \\
NaNbN$_2$       & 166 & 1.1 & 0.8   & 3.70 & 3.3      & -142 & 42     &1.1  & 4 & $-$ \\
HfTaNO$_3$      & 25  & 2.0 & 0.8   & 4.22 & 2.4      & -122 & 73     &1.1  & 0.6 & $-$ \\
KNbO$_3$        & 99  & 1.6 & 3.0   & 1.82 & 1.3      & -367 & 75.5   &1.1  & 3 & $-$ \\
HfSiO$_4$       & 141 & 5.8 & 2.7   & 4.36 & 6.3      & -276 & 18.8   &1.1  & 2 & 190 \\
Y$_2$O$_3$      & 164 & 4.2 & 1.0   & 3.44 & 3.7      & -122 & 41.8   &1.1  & 4 & 5.8 \\
ZrO$_2$         & 225 & 3.7 & 1.1   & 3.78 & 4.6      & -145 & 18.7   &1.1  & 10 & $-$ \\
CdTiO$_3$       & 62  & 2.5 & 1.0   & 5.2  & 4.4     & -205 & 20.7   &0.9  & 2 & $<$1 \\
CaTiO$_3$       & 62  & 2.7 & 1.1   & 5.23 & 3.5     & -170 & 26.8   &0.8  & 1 & $<$1 \\
Y$_2$Ti$_2$O$_7$ & 227 & 3.1 & 1.1  & 2.21 & 10.7     & -224 & 11.4   &0.6  & 2 & $<$1\\                              
\hline
\end{tabular}
\end{ruledtabular}
\end{table*}

We begin our analysis by looking for patterns in the entire dataset.  First, we
note that under the rigid band model used in this work, most materials
have optimal dopings of about $10^{21}$ cm$^{-3}$, which
corresponds to dopings on the order to 10\%.  While this is much
higher than typical semiconductor thermoelectrics, it is consistent
with the behavior of oxides like SrTiO$_3$\cite{srtio3_expt,
  srtio3_expt2,
  oxide_thermo_review1,oxide_thermo_review2,oxide_thermo_review3}. Reaching such high doping values may be difficult in practice, and will require further experimental and theoretical work (see for instance \cite{sns_doping}).  In this work we concentrate on identifying promising materials for further optimization.

In Fig.~\ref{fig:all}, we plot the values of $S$ versus $\sigma/\tau_e$,
at 700K and for a fixed doping of $10^{21}$ cm$^{-3}$.  The
color scale shows the value of $S^2 \sigma / \tau_e$\endnote{We plot
  the direction of $S$ and $\sigma$ which results in the maximum power
  factor}.  There is a clear trade-off between $S$ and $\sigma/\tau_e$,
which is consistent with the behavior of simple parabolic bands where
$S \propto m_{eff}$ and $\sigma \propto 1/m_{eff}$, where $m_{eff}$ is
the effective mass\cite{thermo_review}.  The best materials do not
maximize either $S$ or $\sigma / \tau_e$, but instead have $S$ and
$\sigma / \tau_e$ values in the center of observed range, but with a
larger combination than is typical.  In the following sections, we
explore in more detail how some of these individual materials achieve this
higher than expected combination of $S$ and $\sigma/\tau_e$.

The trade-off between $S$ and $\sigma/\tau_e$ makes finding a simple
descriptor of the power factor in terms of features of the band
structure very difficult, even though we can relate $\sigma$ and $S$
individually to features in the band structure. In table
\ref{tab:corr}, we present Spearman rank correlations between several
thermoelectric properties and various descriptors of the band
structure, and several of these relationships are plotted in
Fig. \ref{fig:comp}.

For example, we find that the smallest value of the effective mass
tensor, $m_{min}$, is highly correlated with $\sigma / \tau_e$, and we
plot this relationship in Fig. \ref{fig:comp}b.  Unsurprisingly,
materials with small effective masses usually have high
conductivities, although this relationship can be complicated by
anisotropy in the effective mass tensor or by many bands contributing
to the conduction. Similarly, as shown in Fig.~\ref{fig:comp}a, we find that we can model
the Seebeck coefficient as $S \propto m_{DOS}(T,n_d)$ (see Eq. \ref{eq:mdos2}).

\begin{table}
\caption{\label{tab:corr} Spearman rank correlation matrix of various
  band structure descriptors and thermoelectric quantities at 700 K
  and at a fixed doping of $10^{21}$ cm$^-3$. $S^2 \sigma / \tau_e$ is
  the power factor, $|S|$ is the absolute value of the Seebeck
  coefficient, $\sigma / \tau_e $ is the electrical conductivity,
  $m_{min}$ is the minimum value of $m_{ij}$, the effective mass
  tensor, $m_{DOS}$ is defined in Eq. \ref{eq:mdos2}, and DOS is the
  density of states at the Fermi level.  Correlations with absolute value above 0.65 are
  in bold.}
\begin{ruledtabular}
\begin{tabular}{lcccccccccc}
Quantity & $S^2 \sigma / \tau_e$ & $|S|$ & $\sigma/\tau_e$ & $m_{min} $&$m_{DOS}$& $DOS$\\
\hline
$S^2 \sigma / \tau_e$  &    $-$ &    0.44 &    0.07 &   -0.03 &    0.22 &    \bf{0.67} \\
$|S|$                &    0.44 &    $-$ &   \bf{-0.81} &    \bf{0.66} &    \bf{0.84} &    0.43 \\
$\sigma/\tau_e$        &    0.07 &   \bf{-0.81} &    $-$ &   \bf{-0.81} &   \bf{-0.83} &   -0.21 \\
$m_{min} $           &   -0.03 &    \bf{0.66} &   \bf{-0.81} &    $-$ &    \bf{0.74} &    0.28 \\
$m_{DOS}$            &    0.22 &    \bf{0.84} &   \bf{-0.83} &    \bf{0.74} &    $-$ &    0.32 \\
$DOS$                &    \bf{0.67} &    0.43 &   -0.21 &    0.28 &    0.32 &    $-$ \\
\end{tabular}
\end{ruledtabular}
\end{table}

Despite these relatively strong relationships for $S$ and
  $\sigma$ in terms of certain definitions of the effective mass, we
  find that the combination of $m_{DOS}^2 / m_{min}$ has only a weak
  correlation with $S^2 \sigma /\tau_e$, as plotted in
  Fig. \ref{fig:comp}d. The problem is that as shown in table
  \ref{tab:corr}, the two definitions of the effective mass are
  strongly correlated with each other, and dividing one by another
  does not produce a useful descriptor. Other quantities like
  $m_{iso}$ have similar problems.  Furthermore, all of the effective
  masses, as well as $S$ and $\sigma / \tau_e$, are strongly
  correlated or anti-correlated with each other, but none are by
  themselves strongly correlated with the power factor. This can be
  understood in part by looking at Fig. \ref{fig:all}, which shows
  that the best materials do not lie at the extreme of either $S$ or
  $\sigma / \tau_e$, but they instead have an atypical relationship
  between $S$ and $\sigma / \tau_e$. Finding a simple descriptor of
  that relationship is difficult when the stronger trend is the
  trade-off between $S$ and $\sigma / \tau_e$.  In addition, in many
  anisotropic materials, the Seebeck coefficient and conductivity are
  not maximized in the same direction, which makes finding a simple descriptor for the maximum power factor more difficult.
Finally, as we will see below, all of our best
materials have unusual band structures with some combination of high
anisotropy, non-parabolic behavior, and multiple bands contributing to
the conduction, and identifying these qualities requires going beyond a
simple effective mass description of parabolic bands.

The best
simple descriptor we found for $S^2 \sigma /\tau_e$ does not include any effective mass, but instead is just the DOS evaluated at
the relevant doping and temperature, as shown in Fig.~\ref{fig:comp}c.
While there is a clear relationship between the DOS and the power factor,
many high DOS materials have low power factors, making a high DOS a
useful design criterion but not a sufficient condition for good thermoelectric
performance.

In the following sub-sections, we will investigate the band structures
of some of the materials in table \ref{tab:results}, in
order to evaluate the mechanisms that allow these particular
materials to minimize the trade-off between $S$ and $\sigma$.  We find that
these materials separate roughly into three classes of materials,
although some materials fall into several classes. In general, the
mechanisms for high power factors consist of combining a large number of flat bands near the
conduction band minimum with at least some highly dispersive
bands. This combination allows a large number of carriers, some of
which are in dispersive bands, increasing $\sigma$, while at the same
time keeping the Fermi level near the conduction band minimum, where $|S|$ is
largest, mitigating the typical trade-off.

\subsection{Symmetry driven degeneracy}

This group of promising thermoelectrics contains materials that have
symmetries (or near symmetries) which cause degeneracies in the
conduction band minimum.  This degeneracy increases the DOS for any given Fermi level, relative to a
material without degeneracies. These degeneracies can be due to a
single degenerate minimum in Brillouin zone, or they can be due to a
band structure with a conduction band minimum which is repeated due to
symmetry.

\begin{figure}
\includegraphics[width=3.2in]{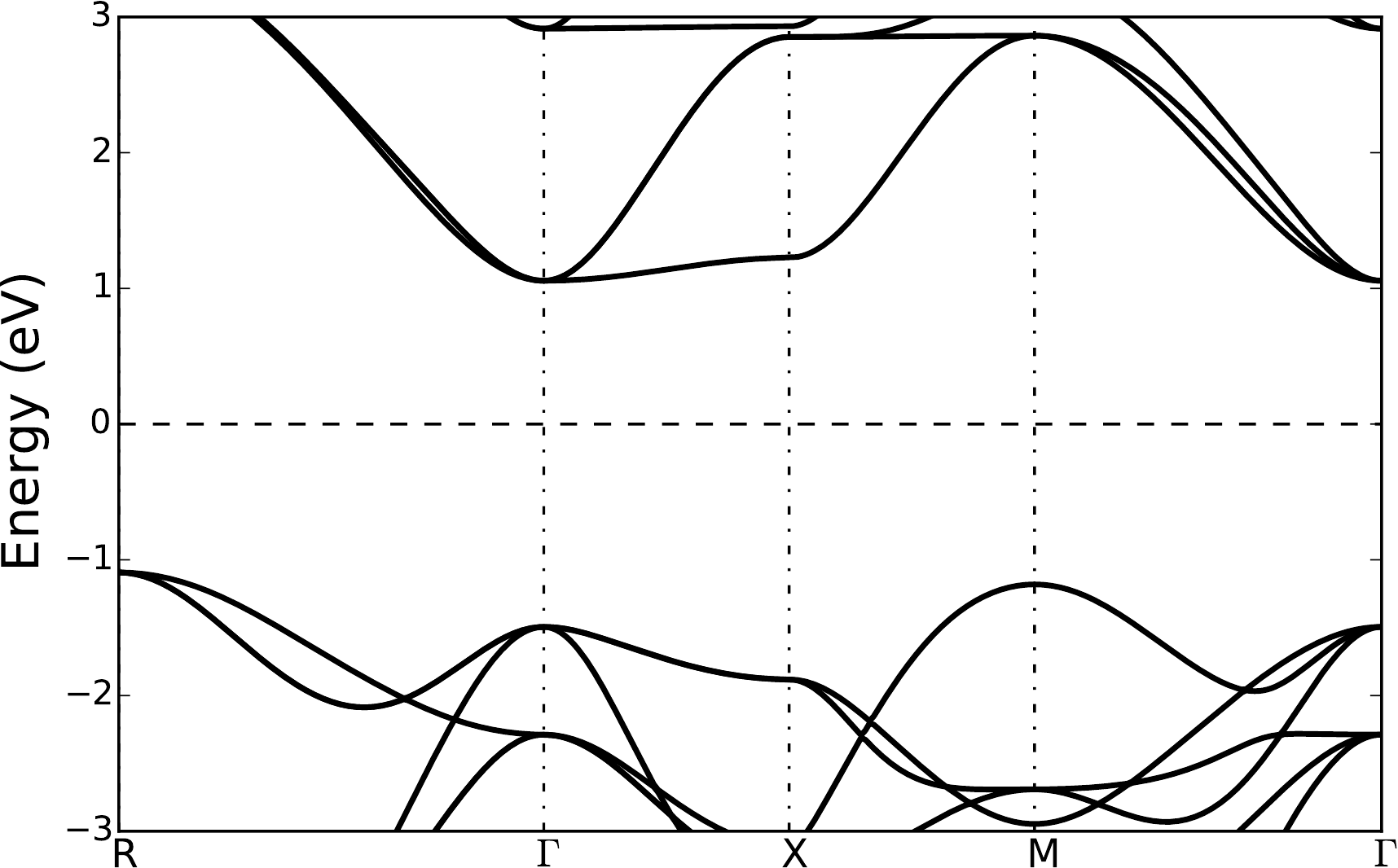}
\caption{\label{fig:srtio3} Band structure of cubic SrTiO$_3$. Energies are relative to the Fermi level.}
\end{figure}

In addition to having degeneracies, the conduction bands of these
materials all consist of empty transition metal $d$-orbitals that
have highly anisotropic dispersions\cite{lowd_bands, non_parabolic,
  srtio3_theory,ktao3_theory}.  These anisotropic band structures
allow the material to have both low $m_{eff}$ and high $m_{eff}$ bands
at the same minimum, combining large Seebeck coefficients
with the high conductivity.  Similar degeneracies and
anisotropic bands are behind the high power factors of several
semiconducting materials which rely of empty $p$-orbitals instead of
empty $d$-orbitals\cite{lowd_bands,non_parabolic,ab_thermoelectrics,thermoelectric_predictors}.

All of these features are present in the band structure of cubic
SrTiO$_3$, as shown in see Fig.~\ref{fig:srtio3}, which is known to be
a good $n$-type thermoelectric.  SrTiO$_3$ has a single triply
degenerate conduction band minimum at $\Gamma$ due to the $t_{2g}$
states originating from the Ti-$d$ orbitals.  These bands have highly
anisotropic dispersions, with one nearly flat band ($m_{eff} =
  6.3$) and two highly dispersive bands ($m_{eff} = 0.4$) going
from $\Gamma$ to $X$. The combination of high degeneracy and high effective mass bands with very
low effective mass bands, which allow for high conductivity, is what allows SrTiO$_3$ to escape the normal trade-off between $S$ and $\sigma / \tau_e$.

Similar features are
present in many of the other perovskite variants which we find to be
candidate thermoelectrics (SrTiO$_3$, PbTiO$_3$, NaNbO$_3$, LiNbO$_3$,
KNbO$_3$, LiTaO$_3$, Ba$_2$TaInO$_6$, CaTiO$_3$, Sr$_2$TaInO$_6$, SrTaNO$_2$).  In addition, various phases of
TiO$_2$ and ZrO$_2$ have similar features which lead to high power
factors.  Many of these materials have been studied as thermoelectrics
before, and the mechanisms leading to their power factors are
relatively well-known\cite{srtio3_theory, ktao3_theory,ab_thermoelectrics,thermoelectric_predictors}, so we will
proceed with a discussion of the next two groups.

\subsection{Low dimensional conductors}

\begin{figure}
\includegraphics[width=3.2in]{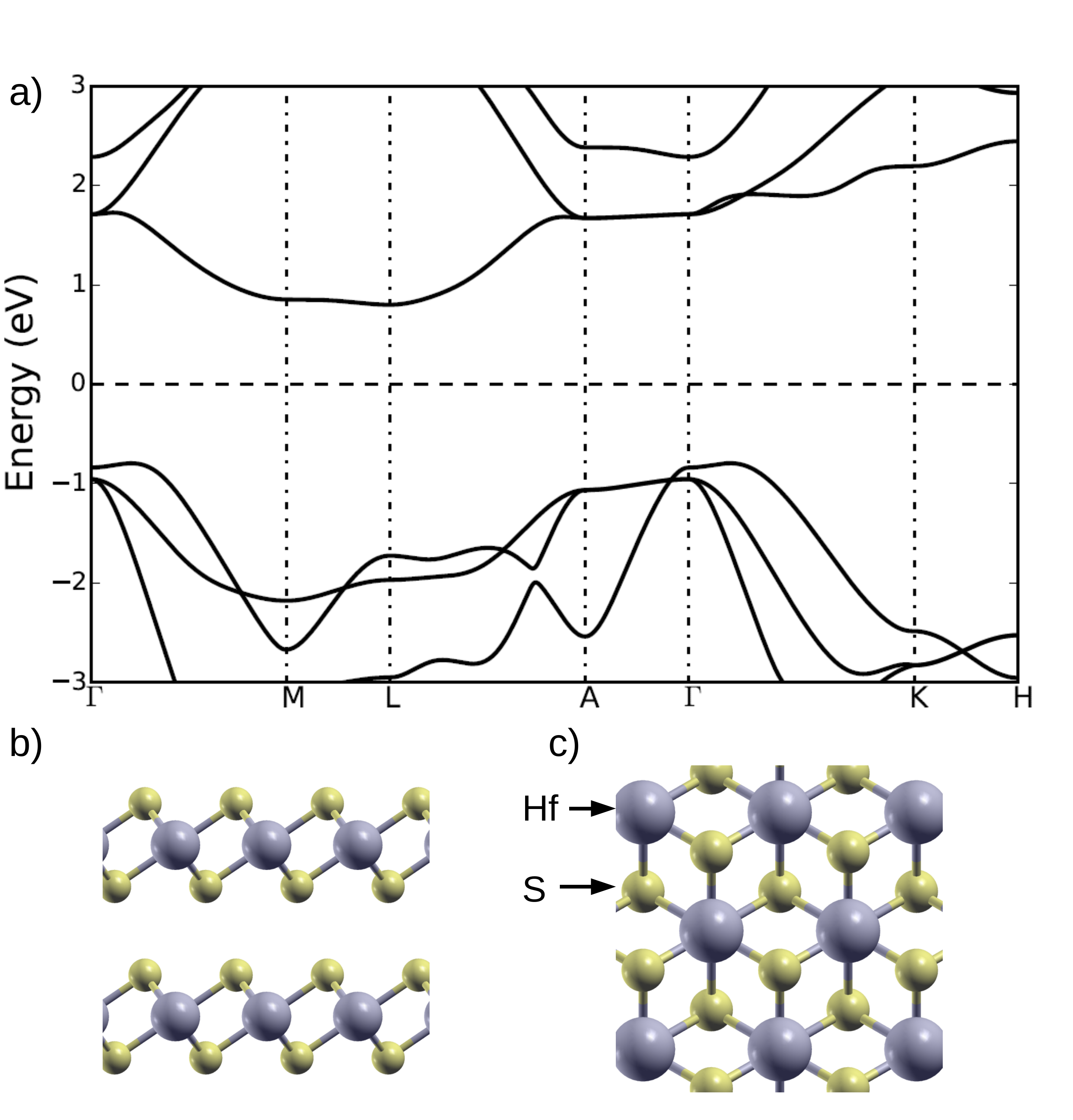}
\caption{\label{fig:hfs2} a) Band structure of HfS$_2$. b-c) Side and top views of HfS$_S$.  Larger gray atoms are Hf, smaller yellow atoms are S.}
\end{figure}

While all of the structures studied in this work are three dimensional,
in many cases the atoms which dominate the conduction band minima are
arranged in two-dimensional layers, one-dimensional lines, or
zero-dimensional dots, which leads to effectively low dimensional
conduction.  In some cases, the material itself
consists of weakly bound layers, while in others
there are strong bonds in all three directions, but the transition
metals are arranged in a low-dimensional way.

Reducing the effective dimensionality of a material results in highly
anisotropic conduction bands and an increased DOS at the bottom of
bands, which can increase the power factor\cite{low_dim_thermo_2, lowd_bands, low_dim_thermo}.  The idea of improving the
power factor of a candidate thermoelectric by reducing its
dimensionality and therefore increasing its DOS is well-known, and has
been shown in SrTiO$_3$ superlattices\cite{srtio3_lowd,
  quantum_well_thermo}. We note that here we are considering
thermodynamically stable materials, rather than artificial
superlattices, nanowires, or quantum dots, which should reduce
manufacturing costs and increase thermodynamic stability.

We present two examples of effectively low dimensional materials which
we predict have high power factors.  First, in Fig.~\ref{fig:hfs2}, we show the
band structure and atomic structure of HfS$_2$, which consists of
weakly bound two-dimensional hexagonal trilayers.  The conduction
bands are very flat from $M$ to the minimum at $L$ ($m_{eff}=4.5$), characteristic of
two-dimensional materials, but they are much more dispersive in other directions ($m_{eff}=0.3$).

\begin{figure}
\includegraphics[width=3.2in]{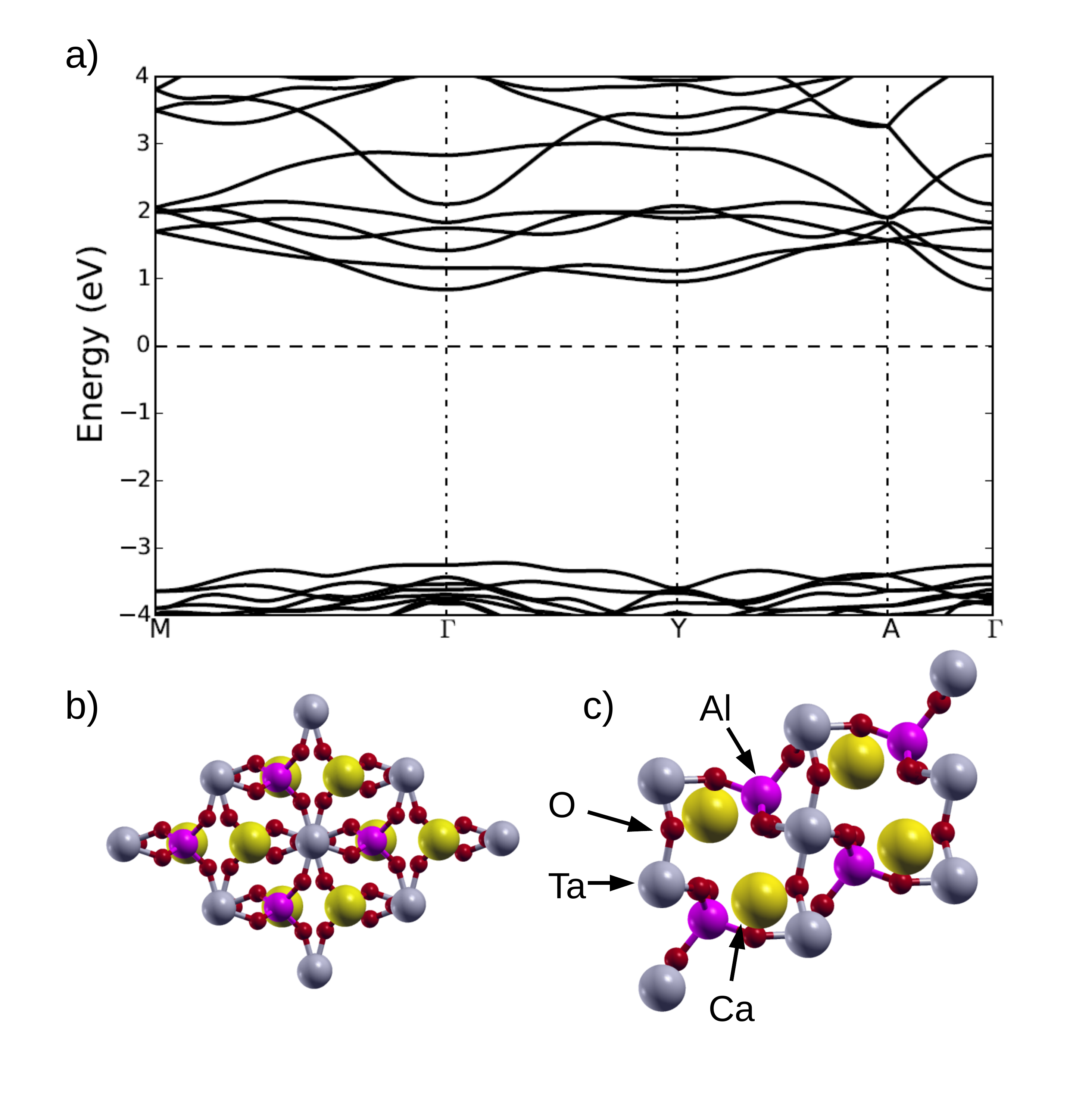}
\caption{\label{fig:cataalo} a) Band structure of CaTaAlO$_5$. b-c) Top
  and side views of CaTaAlO$_5$.  Large yellow atoms are Ca, medium gray atoms
  are Ta, smaller magenta atoms are Al, smallest red atoms are O.}
\end{figure}

Second, in Fig.~\ref{fig:cataalo}, we show the band structure and atomic
structure of CaTaAlO$_5$, which consists of TaO$_6$ octahedra arranged
into one-dimensional columns that are separated from each other by Ca ions and AlO$_4$
tetrahedra.  This arrangement of Ta atoms leads to an anisotropic band
structure with very flat bands from $\Gamma$ to $Y$ ($m_{eff}=2.4$) but stronger
dispersion from $\Gamma$ to $A$ ($m_{eff} =0.5$).  There are additional nearly degenerate conduction band minima at $Y$ which also contribute to the DOS.
In HfS and CaTaAlO$_5$, the high DOS and the 
strong anisotropy, which are caused by the low dimensionality, create the conditions for a high power factor.

Within our set of candidate thermoelectrics, ZrS$_2$, TiS$_2$, HfS$_2$, YClO,
CaTiSiO$_5$, WP$_2$O$_8$, TaPO$_5$, NaNbN$_2$ have quasi-two-dimensional
structures, CaTaAlO$_5$, HgWO$_4$, LaTaO$_4$, and HfSiO$_4$ have
quasi-one-dimensional structures, and NbTl$_3$S$_4$, Ba$_2$TaInO$_6$, and
Sr$_2$TaInO$_6$ have quasi-zero-dimensional structures, as their transition metals are separated from each other.

There are other possible advantages in using low-dimensional materials
as thermoelectrics besides the increased DOS, including potentially
lower thermal conductivity, due to phonon scattering from the atomic layers, 
as well as the ability to physically separate dopants from conducting
channels, which can reduce electron scattering.  One disadvantage is
that the thermoelectric properties of low-dimensional materials will be
anisotropic, resulting in reduced efficiency in polycrystalline samples.

\subsection{Accidental Degeneracies}

\begin{figure}
\includegraphics[width=3.2in]{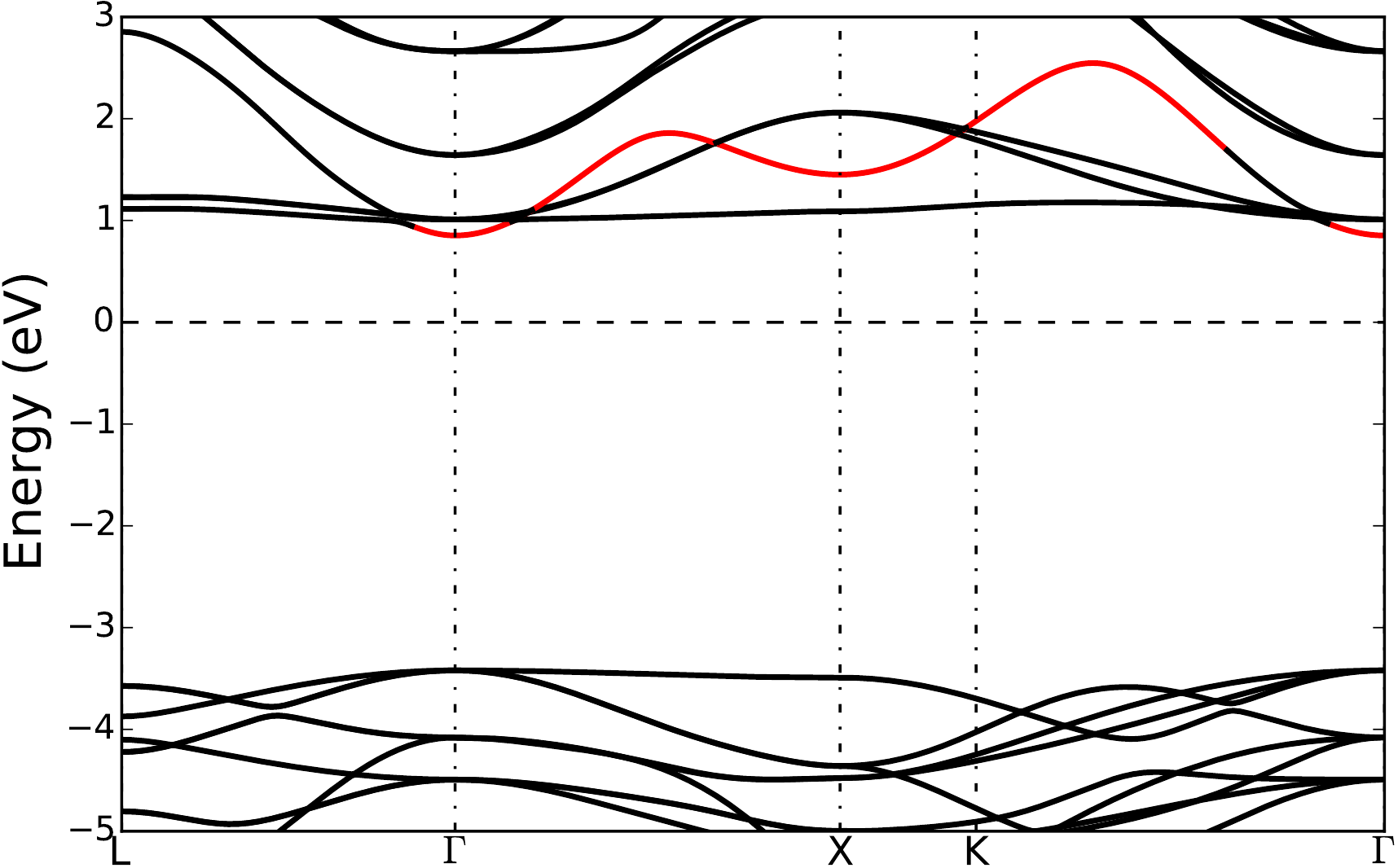}
\caption{\label{fig:srinwo} Band structure of Sr$_2$TaInO$_6$. Bands
  with greater then 35\% In content are colored red, others are
  black.}
\end{figure}

One final mechanism for increasing the power factor of an $n$-type
oxide is to find or engineer a material with accidental degeneracies
of the conduction band minimum.  While this can happen for physically
similar bands which happen to be degenerate at different points in the
Brillouin zone, here we consider cases where the bands come from
different orbitals and have different effective masses.  For example,
in the double perovskite Sr$_2$TaInO$_6$, the conduction band consists
of both Ta $d$-states and In $p$-states, which happen to be at similar
energies (see Fig.~\ref{fig:srinwo}, which highlights the In states in
red).  The In states have low effective masses ($m_{eff} = 0.2-0.3$) while the Ta states
have much higher effective masses ($m_{eff} = 13-62$), allowing the material to take advantage
of both types of bands, in addition to the increased DOS provided by
the near degeneracy.  The large effective masses of the Ta-$d$ bands are caused by small overlap between them and many of the neighboring In-$p$ orbitals,
which results in very flat bands, which contribute
to a high DOS and high power factor.

Similar materials with two different atoms contributing to the conduction
are HgWO$_4$, Ba$_2$TaInO$_6$ and NbTl$_3$S$_4$, where the Hg(+2), In(+3) and Tl(+1)
ions, contribute empty $s/p$-bands at similar energies to the transition
metal $d$-bands.  In addition, in both YClO and Y$_2$O$_3$, the empty $s$
and $d$-states of the Y atoms are both located near the conduction band minimum, which results
in similar behavior to the case where the orbitals come from different atoms.  Depending on the crystal structure and the anions, it
may be possible to engineer empty $s$-bands from Cu, Zn, Ag, Cd, Au, or Hg or
$p$-bands from In, Sn, Tl, Pb, or Bi to become degenerate with transition metal bands in this fashion.  This
type of engineering could allow one material to take advantage of the high Seebeck
coefficients of transition metal oxides while incorporating the higher
mobility of semiconductors, which often have empty $s$ or $p$ orbitals
from main group elements.  The exact alignments of empty
states from different atoms is difficult to predict using DFT+U, so
further study of these materials to determine the band alignments more
precisely may be necessary.

\subsection{Thermal Conductivity}

Due to the high computational cost, we were not able to calculate the
thermal conductivity of our full dataset.  For 191 compounds, we
calculated the Debye temperature (see Eq. \ref{eq:debye}), which is fairly strongly correlated
with thermal conductivity (see table \ref{tab:thermal} and supplementary materials).  Calculations of the Debye
temperature are both less sensitive to the q-point
sampling of the phonon band structure than the Gruneisen parameter and
require the phonons at only one volume, making the computations much faster.

We find that in our set of transition metal oxides, nitrides, and
sulfides, there is relatively little variation in the Debye
temperature (mean of 342 K, standard deviation of 66 K), as compared
to our test dataset of simple binary and ternary semiconductors (mean 319
K, standard deviation 234 K).  This is likely due to the fact that all
of our compounds contain ionic bonds between light anions and medium
to heavy transition metals, while the test dataset contains a range of
bonds, from covalent to ionic, and a range of atom masses. In both
datasets, there is a significant correlation between $V^{-1}$ and
the Debye temperature, with a correlation coefficient of 0.68 in the oxides,
and 0.88 in the test set.  This suggests that looking at oxides with
larger unit cells could be beneficial\cite{thermalcond_trends,thermoelectric_predictors}.

Due to the relatively weak variation in the Debye temperature
throughout our set of oxides, the Gruneisen parameter becomes more
important to identify the materials likely to have low thermal
conductivity. Many of the oxides we consider have soft or unstable
phonon modes, which likely results in strong anharmonicity and low thermal 
conductivity, but this is difficult to quantify without more involved calculations. 
Due to the large computational cost, we are only able to calculate the Gruneisen parameter of the materials in table \ref{tab:results}.
We do not have a large enough database of
oxide Gruneisen parameters to identify any trends which would predict
which materials will have soft modes without doing phonon calculations.

As shown in table~\ref{tab:results}, many of the materials we have
identified as having promising power factors also have low thermal
conductivity according to our model. Most of the perovskite materials we study have strongly
anharmonic modes, which leads to relatively low thermal conductivities,
both in our calculations and in experiment\cite{thermal_perovskites}.
In addition, we find that many of the materials with one or two
dimensional bonding also have soft modes, likely due to the fact that
many of the atoms are relatively free to vibrate in at least one
direction.

\section{\label{conclu}Conclusions}

We have used high throughput first principles calculations to search
for $n$-type transition metal oxides, nitrides, and sulfides which are
promising for thermoelectric applications.  We find many materials
with estimated power factors which are comparable to or surpass
previously studied oxide thermoelectrics.

Across the entire sample of compounds, we find the expected
correlations between the Seebeck coeffient and electrical conductivity
with the effective mass and inverse effective mass, respectively, of
the conduction band electrons.  However, to find materials with high
power factors, it is necessary to look for materials which are not
well described by a single parabolic band, but instead have
degeneracy, anisotropy, or other features which result in a high density of states combined with dispersive bands at the Fermi level. 
These materials achieve their high power factors due to some combination of
symmetry-enforced degeneracies, low dimensionality, or accidental
degeneracies. In addition, we use phonon
calculations to model the thermal conductivity of our best candidates,
and we find many that have low lattice thermal conductivity or that require anharmonic stabilization of the harmonic modes. We hope
further work on these materials, as well as the understanding gained
by examining the mechanisms which lead to high power factors in
oxides, will lead to improved thermoelectric performance in oxides.

\begin{acknowledgments}
We wish to acknowledge discussions with Igor Levin and help with the
ICSD from Vicky Karen and Xiang Li.
\end{acknowledgments}


%
%
%
%
%
%
%



\clearpage

\begin{center}
\widetext{\bf{Supplementary Materials: First principles search for $n$-type oxide, nitride, and sulfide thermoelectrics}}

\end{center}

\setcounter{equation}{0}
\setcounter{figure}{0}
\setcounter{table}{0}
\setcounter{page}{1}
\setcounter{section}{0}

Supplementary materials for ``First principles search for $n$-type oxide, nitride, and sulfide
thermoelectrics.''  Includes larger thermoelectric dataset and data for thermal conductivity testing, as well as some details of Wannier construction.


\section{Wannier States}

Table \ref{tab:wan} contains a list of Wannier states included in the valence by default.

\begin{table}[b]
\caption{\label{tab:wan} List of states to include in Wannierization.}
\tiny
\begin{tabular}{lc}
\hline
\hline
Atom & states \\
\hline
H & $s$\\
He & $s$\\
Li & $s$\\
Be & $s$,$p$\\
B & $s$,$p$\\
C & $s$,$p$\\
N & $p$\\
O & $p$\\
F & $p$\\
Ne & $s$,$p$\\
Na & $s$,$p$\\
Mg & $s$,$p$\\
Al & $s$,$p$\\
Si & $s$,$p$\\
P & $p$\\
S & $p$\\
Cl & $p$\\
Ar & $s$,$p$\\
K & $s$,$d$\\
Ca & $s$,$d$\\
Sc & $s$,$d$\\
Ti & $s$,$d$\\
V & $s$,$d$\\
Cr & $s$,$d$\\
Mn & $s$,$d$\\
Fe & $s$,$d$\\
Co & $s$,$d$\\
Ni & $s$,$d$\\
Cu & $s$,$d$\\
Zn & $s$,$p$,$d$\\
Ga & $s$,$p$\\
Ge & $s$,$p$\\
As & $s$,$p$\\
Se & $s$,$p$\\
Br & $s$,$p$\\
Kr & $s$,$p$\\
Rb & $s$,$d$\\
Sr & $s$,$d$\\
Y & $s$,$d$\\
Zr & $s$,$d$\\
Nb & $s$,$d$\\
Mo & $s$,$d$\\
Tc & $s$,$d$\\
Ru & $s$,$d$\\
Rh & $s$,$d$\\
Pd & $s$,$d$\\
Ag & $s$,$d$\\
Cd & $s$,$p$\\
In & $s$,$p$\\
Sn & $s$,$p$\\
Sb & $s$,$p$\\
Te & $s$,$p$\\
I & $s$,$p$\\
Xe & $s$,$p$\\
Cs & $s$,$d$\\
Ba & $s$,$d$\\
La & $s$,$d$,$f$\\
Hf & $s$,$d$\\
Ta & $s$,$d$\\
W & $s$,$d$\\
Re & $s$,$d$\\
Os & $s$,$d$\\
Ir & $s$,$d$\\
Pt & $s$,$d$\\
Au & $s$,$d$\\
Hg & $s$,$p$,$d$\\
Tl & $s$,$p$\\
Pb & $s$,$p$\\
Bi & $s$,$p$\\
Po & $s$,$p$\\
At & $s$,$p$\\
\hline
\hline
\end{tabular}
\end{table}

\section{Thermal Conductivity}

Table \ref{tab:thermal} contains the testing of our thermal conductivity method.
$\kappa_l^{Bjerg} $ uses the original method of Bjerg
\textit{et. al.}\cite{model_thermal}.
$\kappa_l^{Slack}$ uses the
method used in the main text (see Eqs. 2-5 in main text).  The Debye temperature and Gruneisen parameter follow the definition of Bjerg (Eq. 2-4 in main text), but with
the Slack formula to combine them (Eq. 5 in main text), times 0.7, an empirical correction.
The reference thermal conductivities are from the experimental and
first principles thermal conductivities collected in
Ref. \onlinecite{ht_thermal_cond}, largely from Refs. \onlinecite{hh_thermal,slackbook}, see references therein.

\begin{table*}
\caption{\label{tab:thermal} Thermal conductivities at 300 K. hh stands for half-Heusler.}
\begin{ruledtabular}
\tiny
\begin{tabular}{lccccccc}
Material & Struct. & Vol      & Debye Temp. & Gruneisen & $\kappa_l^{Bjerg} $ &$\kappa_l^{Slack} $ & $\kappa_l^{Reference}$ \\
         &         & Bohr$^3$ & (K)         &          & W/(m K)& W/(m K)& W/(m K) \\
\hline
KAgTe & hh & 583 & 138 & 5.2 & 0.3 & 0.3 & 0.5 \\
BaLiSb & hh & 613 & 226 & 2 & 6.1 & 8.2 & 0.6 \\
AgCl & rocksalt & 272 & 167 & 3.4 & 1.1 & 0.6 & 1 \\
SrPdTe & hh & 530 & 137 & 3.4 & 0.5 & 0.7 & 1.2 \\
SnTe & rocksalt & 418 & 124 & 6.3 & 1.5 & 0.1 & 1.5 \\
AgTeLi & hh & 467 & 213 & 1.9 & 2.7 & 6.2 & 1.5 \\
\end{tabular}
\end{ruledtabular}
\end{table*}

\section{All results}

Table \ref{tab:seeb} contains our large dataset of results.  The table
is labeled like the main table in the text.  The extra columns are the
ICSD entry number for beginning the relaxation, the volume, and near
the end, the Debye temperature in Kelvin and the maximal $S^2 \sigma/\tau$ at 700 K, but for $p$-type
doping.

\begin{table*}
\caption{\label{tab:seeb} Full results table.}
\begin{ruledtabular}
\tiny
\begin{tabular}{lcccccccccccccc}
Material & Space Grp. & ICSD & Vol.    & Band Gap & $m_{eff}$ & DOS                            & $m_{DOS}$ & $S$           & $ \sigma/\tau$                           &  $S^2 \sigma/\tau$                          & Opt. Doping         & Debye T & $S^2 \sigma/\tau$  \\
         &            &       & \AA$^3$ &     (eV) &           & ($10^3$ eV$^{-1} $\AA$^{-3}) $ &           & ($\mu V / K$) & 10$^{-3}$ $\frac{\text{S}}{\text{m fs}}$ & 10$^3$ $\frac{W}{\text{m K}^{2} \text{fs}}$ & ($10^{21} cm^{-3}$) & K & $p$-type            \\
\hline
NaCoO$_{2}$ & 166 & 96428 & 76.0 & 1.9 & $-$ &  $-$  & $-$ & -175.1 & 35.3 & 2.47 & 9.0 & $-$ & 1.53 \\
FNTi & 59 & 290034 & 73.9 & 1.0 & 4.8 & 3.3 & 3.9 & -386.1 & 36.4 & 2.25 & 5.0 & 420.2 & 0.64 \\
AlCaO$_{5}$Ta & 15 & 50718 & 200.9 & 4.0 & 1.2 & 4.1 & 5.0 & -189.1 & 42.9 & 1.74 & 3.0 & 340.2 & 0.37 \\
O$_{2}$Ti & 225 & 189325 & 28.0 & 1.7 & 1.6 & 1.5 & 17.0 & -305.3 & 9.6 & 1.68 & 8.0 & 570.1 & 0.98 \\
LiNbO$_{3}$ & 161 & 28300 & 104.8 & 3.5 & 1.3 & 3.9 & 7.0 & -244.7 & 23.4 & 1.56 & 2.0 & 348.4 & 0.75 \\
O$_{2}$Ti & 136 & 85494 & 63.4 & 2.1 & 1.0 & 3.6 & 7.5 & -250.3 & 20.7 & 1.56 & 3.0 & 411.4 & 0.43 \\
\end{tabular}
\end{ruledtabular}
\end{table*}
\begin{table*}
\caption{Full results table confinued}
\begin{ruledtabular}
\tiny
\begin{tabular}{lcccccccccccccc}
Material & Space Grp. & ICSD & Vol.    & Band Gap & $m_{eff}$ & DOS                            & $m_{DOS}$ & $S$           & $ \sigma/\tau$                           &  $S^2 \sigma/\tau$                          & Opt. Doping         & Debye T & $S^2 \sigma/\tau$  \\
         &            &       & \AA$^3$ &     (eV) &           & ($10^3$ eV$^{-1} $\AA$^{-3}) $ &           & ($\mu V / K$) & 10$^{-3}$ $\frac{\text{S}}{\text{m fs}}$ & 10$^3$ $\frac{W}{\text{m K}^{2} \text{fs}}$ & ($10^{21} cm^{-3}$) & K & $p$-type            \\
\hline

InNbO$_{4}$ & 13 & 100695 & 144.3 & 3.3 & 1.2 & 4.6 & 4.8 & -201.5 & 18.3 & 0.80 & 2.0 & $-$ & 0.67 \\
NOTa & 12 & 173006 & 160.2 & 2.4 & 1.5 & 4.9 & 2.8 & -171.5 & 37.5 & 0.80 & 0.8 & 370.2 & 0.37 \\
BaO$_{3}$Ti & 123 & 31885 & 63.8 & 2.1 & 1.1 & 5.0 & 3.0 & -158.5 & 31.8 & 0.80 & 1.0 & 377.8 & 0.96 \\
O$_{2}$Zr & 137 & 92093 & 66.8 & 4.2 & 1.0 & 2.6 & 6.1 & -122.2 & 22.7 & 0.80 & 9.0 & 422.2 & 3.14 \\
BaO$_{3}$Ti & 221 & 186437 & 63.8 & 2.1 & 1.1 & 5.0 & 3.0 & -158.6 & 31.7 & 0.80 & 1.0 & 377.8 & 0.96 \\
FeO$_{3}$Ti & 62 & 246514 & 201.9 & 0.7 & $-$ &  $-$  & $-$ & -196.6 & 17.4 & 0.79 & 3.0 & $-$ & 0.14 \\
\end{tabular}
\end{ruledtabular}
\end{table*}
\begin{table*}
\caption{Full results table confinued}
\begin{ruledtabular}
\tiny
\begin{tabular}{lcccccccccccccc}
Material & Space Grp. & ICSD & Vol.    & Band Gap & $m_{eff}$ & DOS                            & $m_{DOS}$ & $S$           & $ \sigma/\tau$                           &  $S^2 \sigma/\tau$                          & Opt. Doping         & Debye T & $S^2 \sigma/\tau$  \\
         &            &       & \AA$^3$ &     (eV) &           & ($10^3$ eV$^{-1} $\AA$^{-3}) $ &           & ($\mu V / K$) & 10$^{-3}$ $\frac{\text{S}}{\text{m fs}}$ & 10$^3$ $\frac{W}{\text{m K}^{2} \text{fs}}$ & ($10^{21} cm^{-3}$) & K & $p$-type            \\
\hline

BrNZr & 164 & 25507 & 128.1 & 2.8 & 1.5 & 3.8 & 2.6 & -313.6 & 41.9 & 0.61 & 0.5 & $-$ & 0.96 \\
O$_{2}$Ti & 205 & 189326 & 116.4 & 1.8 & 0.8 & 4.4 & 5.9 & -179.4 & 17.6 & 0.61 & 3.0 & $-$ & 0.80 \\
LaMgNaO$_{6}$W & 11 & 40497 & 243.9 & 2.9 & 1.9 & 3.0 & 3.6 & -411.1 & 22.9 & 0.60 & 4.0 & $-$ & 0.57 \\
FOY & 166 & 30623 & 80.5 & 5.2 & 0.3 & 5.2 & 2.7 & -137.8 & 36.1 & 0.60 & 0.9 & $-$ & 0.47 \\
HfO$_{2}$ & 61 & 173965 & 272.4 & 4.7 & 1.3 & 5.2 & 3.7 & -166.4 & 21.7 & 0.60 & 1.0 & $-$ & 0.85 \\
O$_{5}$Ta$_{2}$ & 5 & 280397 & 76.8 & 2.5 & 1.2 & 4.3 & 3.7 & -167.7 & 23.0 & 0.60 & 2.0 & $-$ & 0.98 \\
\end{tabular}
\end{ruledtabular}
\end{table*}
\begin{table*}
\caption{Full results table confinued}
\begin{ruledtabular}
\tiny
\begin{tabular}{lcccccccccccccc}
Material & Space Grp. & ICSD & Vol.    & Band Gap & $m_{eff}$ & DOS                            & $m_{DOS}$ & $S$           & $ \sigma/\tau$                           &  $S^2 \sigma/\tau$                          & Opt. Doping         & Debye T & $S^2 \sigma/\tau$  \\
         &            &       & \AA$^3$ &     (eV) &           & ($10^3$ eV$^{-1} $\AA$^{-3}) $ &           & ($\mu V / K$) & 10$^{-3}$ $\frac{\text{S}}{\text{m fs}}$ & 10$^3$ $\frac{W}{\text{m K}^{2} \text{fs}}$ & ($10^{21} cm^{-3}$) & K & $p$-type            \\
\hline

Ba$_{2}$O$_{6}$SrW & 148 & 246114 & 156.2 & 3.1 & 1.0 & 4.1 & 3.1 & -128.3 & 27.1 & 0.45 & 0.8 & $-$ & 0.44 \\
BaHfO$_{8}$P$_{2}$ & 12 & 245690 & 180.9 & 5.2 & 0.7 & 4.5 & 3.0 & -129.2 & 26.8 & 0.45 & 0.9 & $-$ & 0.44 \\
ClNZr & 166 & 151468 & 106.9 & 1.9 & 3.3 & 2.5 & 1.0 & -79.4 & 49.8 & 0.45 & 0.3 & $-$ & 0.32 \\
Ba$_{2}$BiO$_{6}$Ta & 12 & 153122 & 155.9 & 2.4 & 0.5 & 3.4 & 3.1 & -118.3 & 36.5 & 0.45 & 2.0 & $-$ & 0.41 \\
Fe$_{2}$K$_{2}$Nb$_{4}$O$_{13}$ & 12 & 50038 & 290.8 & 1.4 & $-$ &  $-$  & $-$ & -97.3 & 33.5 & 0.45 & 0.4 & $-$ & 0.27 \\
HgO$_{4}$W & 64 & 169671 & 286.0 & 1.1 & 0.9 & 2.9 & 1.4 & -75.2 & 59.8 & 0.45 & 0.4 & $-$ & 0.34 \\
\end{tabular}
\end{ruledtabular}
\end{table*}
\begin{table*}
\caption{Full results table confinued}
\begin{ruledtabular}
\tiny
\begin{tabular}{lcccccccccccccc}
Material & Space Grp. & ICSD & Vol.    & Band Gap & $m_{eff}$ & DOS                            & $m_{DOS}$ & $S$           & $ \sigma/\tau$                           &  $S^2 \sigma/\tau$                          & Opt. Doping         & Debye T & $S^2 \sigma/\tau$  \\
         &            &       & \AA$^3$ &     (eV) &           & ($10^3$ eV$^{-1} $\AA$^{-3}) $ &           & ($\mu V / K$) & 10$^{-3}$ $\frac{\text{S}}{\text{m fs}}$ & 10$^3$ $\frac{W}{\text{m K}^{2} \text{fs}}$ & ($10^{21} cm^{-3}$) & K & $p$-type            \\
\hline
O$_{4}$SrW & 15 & 155794 & 173.1 & 4.4 & 3.9 & 3.3 & 8.9 & -231.2 & 6.5 & 0.35 & 2.0 & $-$ & 0.47 \\
O$_{4}$SrW & 88 & 155425 & 172.9 & 4.5 & 4.0 & 3.2 & 8.8 & -234.1 & 6.3 & 0.35 & 2.0 & $-$ & 0.46 \\
AlO$_{3}$Y & 194 & 27100 & 119.0 & 3.5 & 0.4 & 1.3 & 0.8 & -35.0 & 52.2 & 0.35 & 8.0 & $-$ & 0.64 \\
Li$_{2}$O$_{3}$Ti & 15 & 15150 & 213.0 & 3.4 & 1.9 & 3.2 & 9.9 & -215.3 & 7.7 & 0.34 & 2.0 & $-$ & 0.73 \\
LaO$_{7}$RbTa$_{2}$ & 123 & 81870 & 169.1 & 2.6 & 3.6 & 1.8 & 1.2 & -81.7 & 58.7 & 0.34 & 0.2 & $-$ & 1.58 \\
Ca$_{2}$Nb$_{2}$O$_{7}$ & 227 & 22411 & 280.6 & 1.7 & 0.6 & 2.9 & 7.0 & -138.7 & 15.8 & 0.34 & 3.0 & $-$ & 0.26 \\
\end{tabular}
\end{ruledtabular}
\end{table*}
\begin{table*}
\caption{Full results table confinued}
\begin{ruledtabular}
\tiny
\begin{tabular}{lcccccccccccccc}
Material & Space Grp. & ICSD & Vol.    & Band Gap & $m_{eff}$ & DOS                            & $m_{DOS}$ & $S$           & $ \sigma/\tau$                           &  $S^2 \sigma/\tau$                          & Opt. Doping         & Debye T & $S^2 \sigma/\tau$  \\
         &            &       & \AA$^3$ &     (eV) &           & ($10^3$ eV$^{-1} $\AA$^{-3}) $ &           & ($\mu V / K$) & 10$^{-3}$ $\frac{\text{S}}{\text{m fs}}$ & 10$^3$ $\frac{W}{\text{m K}^{2} \text{fs}}$ & ($10^{21} cm^{-3}$) & K & $p$-type            \\
\hline
CdCu$_{3}$O$_{12}$Ti$_{4}$ & 204 & 39467 & 203.9 & 0.9 & $-$ &  $-$  & $-$ & -129.9 & 12.5 & 0.22 & 0.6 & $-$ & 0.26 \\
InMo$_{2}$O$_{8}$Rb & 164 & 10186 & 225.4 & 4.2 & 3.2 & 4.4 & 8.4 & -277.8 & 5.3 & 0.22 & 1.0 & $-$ & 0.55 \\
K$_{2}$La$_{2}$O$_{10}$Ti$_{3}$ & 139 & 74193 & 225.9 & 2.4 & 7.6 & 1.9 & 1.6 & -409.9 & 37.1 & 0.22 & 8.0 & $-$ & 0.58 \\
La$_{2}$Li$_{2}$O$_{10}$Ti$_{3}$ & 139 & 82907 & 198.2 & 2.4 & 4.3 & 2.5 & 1.6 & -173.4 & 40.2 & 0.21 & 9.0 & 292.3 & 0.83 \\
CO$_{7}$SiY$_{2}$ & 2 & 88878 & 256.1 & 4.9 & 2.6 & 5.5 & 4.1 & -162.5 & 8.7 & 0.21 & 0.6 & $-$ & 0.10 \\
O$_{4}$PbW & 64 & 155522 & 294.2 & 1.5 & 0.8 & 3.0 & 4.3 & -124.1 & 23.3 & 0.21 & 3.0 & $-$ & 0.44 \\
\end{tabular}
\end{ruledtabular}
\end{table*}


%


\end{document}